# Integration of quantum key distribution and gigabit-capable passive optical network based on wavelength-division multiplexing

Wei Sun,[1,a)] Liu-Jun Wang,[1,a)] Xiang-Xiang Sun,[1] Hua-Lei Yin,[1] Bi-Xiao Wang,[1] Teng-Yun Chen,[1,2,b)] and Jian-Wei Pan[1,2)]

[1]*Hefei National Laboratory for Physical Sciences at Microscale and Department of Modern Physics, University of Science and Technology of China, Hefei, Anhui 230026, China*

[2]*CAS Center for Excellence and Synergetic Innovation Center in Quantum Information and Quantum Physics, University of Science and Technology of China, Hefei, Anhui 230026, China*

Classical optical communications may be still the main communications technology for the foreseeable future, so integration of the quantum communication network with existing classical optical communication network is necessary because existing telecommunications infrastructure will be shared. This means multiplexing of quantum key distribution (QKD) and strong classical data signals, delivering quantum signals and classic signals in one fiber. Optical splitters are employed to access each user in a gigabit-capable passive optical network (GPON). In a 4-user network the splitter adds at least 6 dB of optical loss to the quantum channel, a 64-user network the splitter adds 18 dB of optical loss to the quantum channel. The optical splitters restrict the transmission distance and performance of QKD. We propose a new integration program of QKD and GPON based on wavelength-division multiplexing (WDM). At the optical splitting point, we use filters to separate the quantum signals and bypass the optical splitter, avoiding losses produced by the optical splitters. This increases the counting rate of the quantum signals states and the signal to noise ratio (SNR) improves, so a higher key generation rate and a longer transmission distance can be obtained with QKD.

Despite the existence of eavesdroppers, a quantum key distribution (QKD) guarantees sharing of a secret key by two remote parties using information security theory based on quantum mechanics[1-3]. QKD has been rapidly progressing towards practical implementation. To date, secure key distribution has been possible over fiber distances of up to 300 km, and in many cases key rates at the Mbps-level have been achieved using point-to-point links[4-10]. Research on quantum communication networks based on QKD has also progressed[11-13].

Nonlinear noise generated by Raman scattering from strong classical telecommunication channels can interfere with weak quantum signals[14-15]. This noise can affect the performance of QKD. Therefore, current experimental and test networks use dark fibers to transmit quantum signals. This will cause the number of fibers needed to increase as the number of users increases in future quantum networks. The costs of installing or leasing extra dark fibers is high and fiber resources are lacking in some regions, which is an obstacle to future construction of a quantum communication network.

Classical optical communications may be the main communications technology for the foreseeable future, so integration of the quantum communication network with the existing classical optical communication network is needed.

This means multiplexing QKD and strong classical data signals to deliver quantum signals and classic signals in one fiber.The first wavelength-division multiplexing (WDM) scheme for QKD and classical channels was proposed by Townsend in 1997[16]. Other solutions have also been proposed[15-30].

A gigabit-capable passive optical network (GPON) is a characteristic point to multi-point single-fiber bidirectional passive optical network, mostly used in access networks. GPONs will play a vital role in scaling up the number of users in QKD networks because GPON and QKD integration has been proposed. WDM of classic optical communications based on Fabry-Pérot (FP) lasers and QKD has been proposed[29]. A dual feeder fiber structure was proposed to reduce the effect of noise generated by inverse Raman signals from the downstream light in GPONs on QKD, demonstrating coexistence of multi-user QKD and full power data traffic from the GPON[30]. It was shown that smooth integration of QKD in optical metro networks was possible using classical and quantum channels and filtering the time and wavelength domains to reduce the noise[31].

Optical splitters are employed to access each user in the GPON. The bigger the optical splitting ratio (the number of users), the bigger the loss produced. In a 4-user network the splitter adds at least 6 dB of extra optical loss to the quantum channel, a 64-user network the splitter adds at least 18dB of extra optical loss to the quantum channel. To obtain an ideal key generation rate, the link loss of the QKD system should be about 16–20 dB, equivalent to an 80 km standard fiber. A typical fiber link length of the GPON is 20 km, the fiber loss is 0.25 dB/km, the link loss is about 5dB, and the total WDM devices insertion loss of Alice and Bob is about 1 dB. When the optical splitting ratio is 32, the splitter adds at least 15 dB of optical loss to the quantum channel. Because of the large losses, the QKD system cannot run even without taking into account noise produced by the classical channel. The effect of the Raman noise produced by the classical channel on QKD is important.

We propose a new integration program of QKD and GPON based on WDM. At the optical splitting point, we use filters to separate the quantum signals and make the quantum signals bypass the optical splitter, avoiding the losses produced by the optical splitters. This prevents the optical splitters from restricting the transmission distance and performance of QKD.

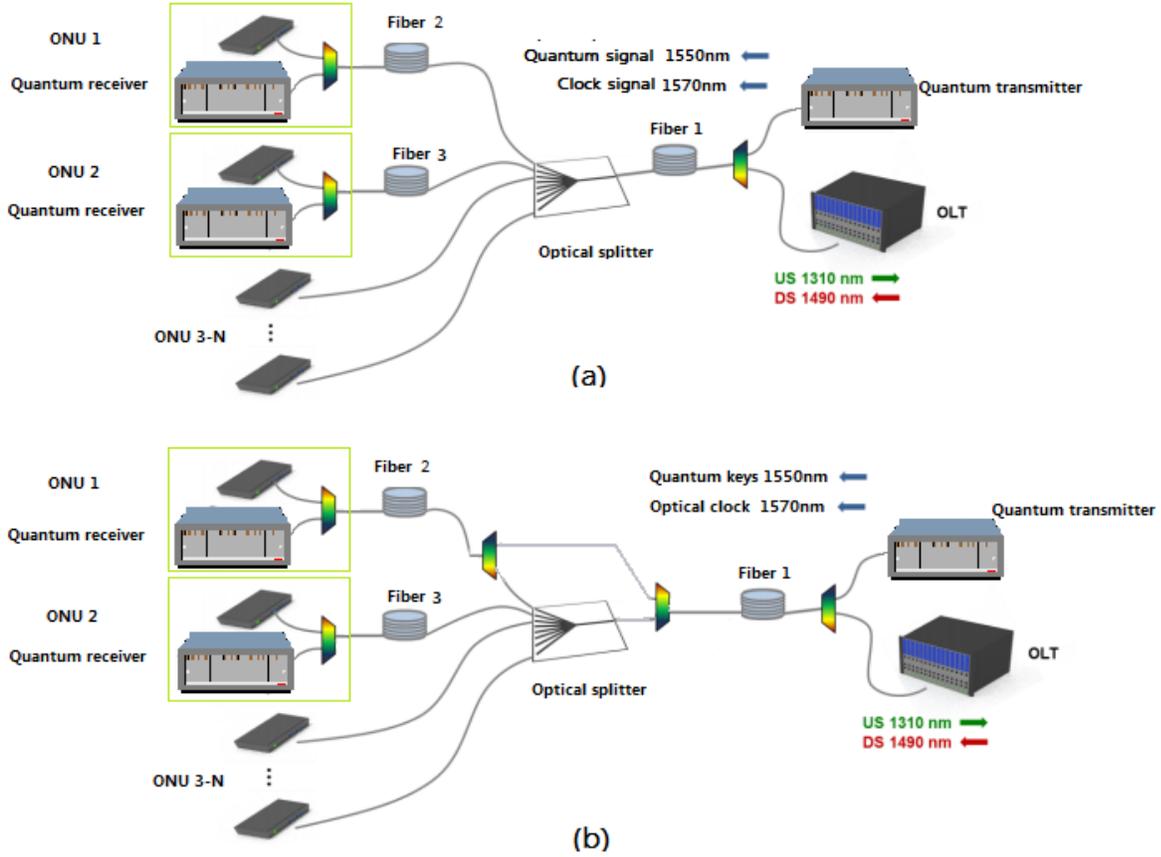

FIG. 1. (a) A schematic illustration of the GPON and QKD WDM: quantum signals and classic signals both pass through the optical splitter. (b) The proposed GPON and QKD WDM: the quantum signals bypass the optical splitter. The red arrows represent the downstream classical signals, the green arrows represent the upstream classical signals, and the blue arrows represent the quantum signals and clock represents the synchronous signals.

The proposed program has the following advantages:

First, the counting rate of the signal state increases N-fold, where N is the optical splitting ratio. At the optical splitting point, the quantum signals bypass the optical splitter, avoiding the loss produced by the optical splitters. This only introduces insertion loss of the 2 WDM device, compared with the optical loss produced by the optical splitter. This loss can be ignored, so the counting rate of the signal state increases N-fold.

Second, the signal to noise ratio (SNR) is improved. When the quantum signals pass through the optical splitter, the forward Raman noise generated by the 1490 nm downstream light from optical line terminal (OLT) in Fiber 1(see Fig. 1) is $d_1$, the inverse Raman noise in Fiber 2 (see Fig. 1) is $d_2$, backward Raman noise generated by the 1310 nm upstream light from optical network unit (ONU) in Fiber 2 is $d_3$, and the forward Raman noise in Fiber 1 $d_4$. The dark count of the detector is $d_0$ and the counting of the signal state is Q. In that case, the SNR is given by:

$$SNR_1 = \frac{Q}{d_0 + d_1 + d_2 + d_3 + d_4}, \qquad (1)$$

When the quantum signals bypass the optical splitter under the same conditions, the inverse Raman noise generated by the 1490 nm downstream signals from OLT in Fiber 1 is given by $Nd_1$, the inverse Raman noise in Fiber 3 is $d_2$, the forward

Raman noise generated by the 1310 nm upstream signals from ONU in Fiber 3 is $d_3$, the forward Raman noise in Fiber 1 is $Nd_4$, and the counting of the signal state changes to NQ. The SNR is then:

$$SNR_2 = \frac{NQ}{d_0 + Nd_1 + d_2 + d_3 + Nd_4}, \qquad (2)$$

The multiplier, K, of the SNR is given by:

$$K = \frac{SNR_2}{SNR_1} = \frac{N(d_0 + d_1 + d_2 + d_3 + d_4)}{d_0 + Nd_1 + d_2 + d_3 + Nd_4} = \frac{N + N\dfrac{d_1 + d_4}{d_0 + d_2 + d_3}}{1 + N\dfrac{d_1 + d_4}{d_0 + d_2 + d_3}}, \qquad (3)$$

As the optical splitting ratio N and the Fiber 1 length changes, the multiplier K changes from 1 to N.

The fiber length from OLT to the optical splitting point is significantly longer than the ONU to the optical splitting point. So, we tested the Raman noise when the OLT (ONU) is placed on Alice's (Bob's) site and when the ONU (OLT) is placed on Bob's (Alice's) site. We found that the Raman noise when the OLT (ONU) is placed on Alice's (Bob's) site is much smaller than when the OLT and Bob were put together so the QKD was better.

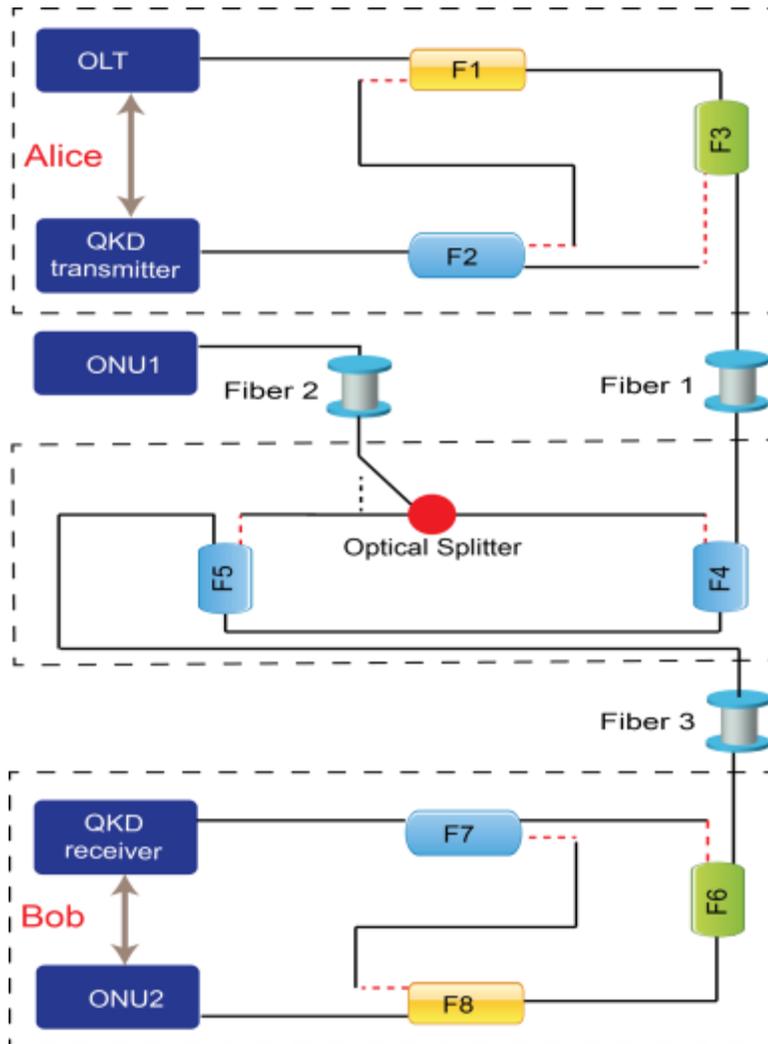

FIG. 2. A new integration program of QKD and GPON based on WDM. The yellow devices F1 and F8 are 1310 nm/1490 nm FWDM, the blue devices F2, F4, F5, and F7 are 1310 nm/1490 nm FWDM. The FWDM is the filter-based wavelength-division multiplexer. The green

devices F3 and F6 are 1490 nm coarse wavelength-division multiplexing (CWDM). The common port of each WDM filter is marked with a short dotted line. The reflect port is on same side as the common-port, while the pass port is on the other side. The central wavelength for F3 and F6 is 1490 nm.

We show the experimental setup in Fig. 2. The wavelengths of the quantum and clock synchronization signals are 1550 nm and 1570 nm. For classical optical communication, we used GPON equipment from Fiberhome, located in Wuhan, China. On Alice's side, F1 is used to clean up and multiplex the classical wavelengths, and F2 and F3 are used to clean up and multiplex the classical and quantum wavelengths. F2 is also used to isolate the 1550 nm background fluorescence in classic signals, which has the center wavelength of 1310 nm, to reduce interference to the quantum signals. At the optical splitting point, F4 and F5 are used to clean up and multiplex the classical and quantum wavelengths. To verify that the GPONs can run, we lead 5 km fiber at the optical splitting point, to give access to user 1 on Bob's side and give access to user 2 to establish communication between user 1 and 2. On Bob's side, F6 and F7 are used to clean up and multiplex the classical and quantum wavelengths and F8 is used to clean up and multiplex the classical wavelengths.

When more filters are added at either Alice's site or Bob's site, they produce more losses than in isolation. This implies that linear crosstalk is eliminated. Our QKD system implements the BB84 polarization coding scheme combined with the decoy method[32-35]. The average photon numbers of the signal and decoy states were chosen to be 0.6 and 0.2, respectively. Vacuum states were also adopted as decoy states. Alice launched three types of states with a ratio of 6:1:1. The system operates at 625 MHz, the single photon detector door width full width at half maxima is 180 ps, and the spectrum bandwidth of Raman noise is limited to 100 GHz in the receiver.

Based on the existing GPON layout, we tested the SNR with the optical splitting ratio varying from 1:4 to 1:128, where the length of fiber 1 + fiber 2 are 12km+2km, 15km+2km, 20km+2km and 12km+12km, respectively. Fiber 1 is the fiber length from Alice (OLT) to the optical splitter point and fiber 2 is the fiber length from Bob (OUN) to the optical splitter point. The QKD index is measured when both QKD and GPON are working at the same time.

Fig. 3 displays the SNR using two different methods with various fiber lengths and different splitting ratios.

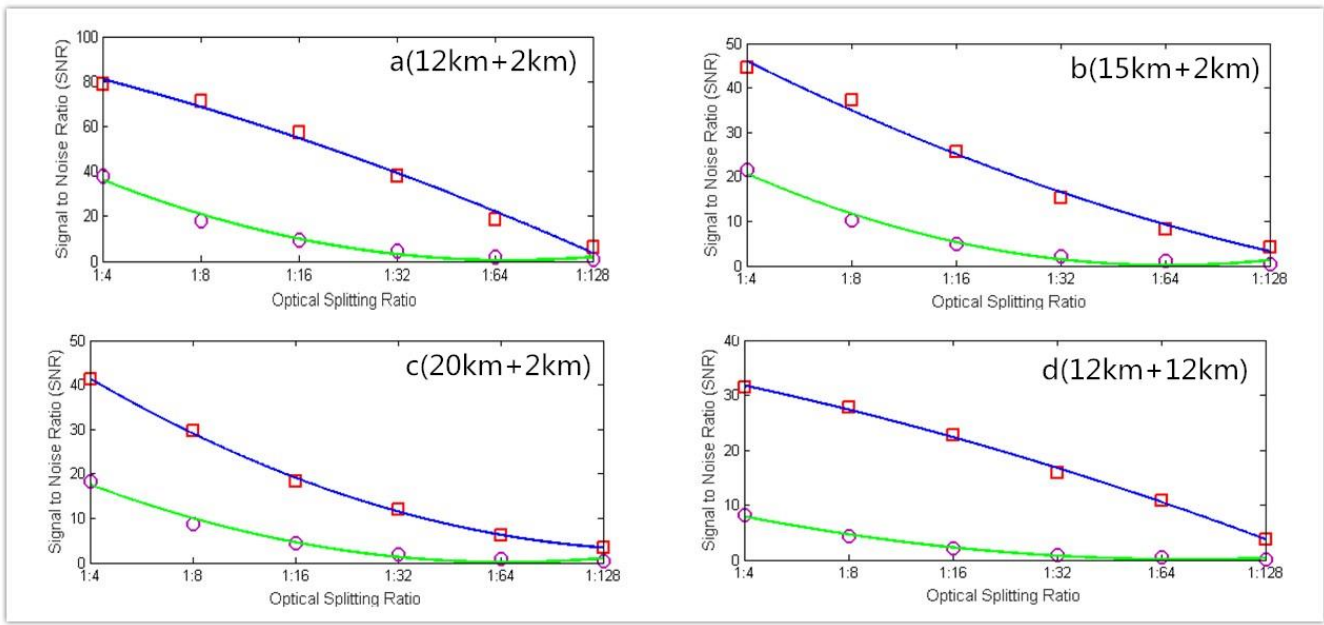

FIG. 3. The SNR measured using two different methods. The blue solid line is the SNR when the quantum signal bypasses the optical splitting point, and the green solid line gives the SNR when the quantum signal goes through the optical splitting point. The fiber length, fiber 1 + fiber 2, from (a) to (d) are 12km+2km, 15km+2km, 20km+2km, and 12km+12km, respectively.

The SNR improves significantly using the new integration methods where the quantum signals bypass the optical splitting point, compared with the traditional method where the quantum signals go through directly. Table I gives the multiplier of the SNR when the quantum signals bypass the optical splitting point.

Table I.. Enhanced multiplier of the SNR of different fiber lengths and different optical splitting ratios.

| Fiber Length | Optical splitting Ratio | | | | | |
|---|---|---|---|---|---|---|
| | 1:4 | 1:8 | 1:16 | 1:32 | 1:64 | 1:128 |
| 12km+2km | 2.09 | 3.89 | 6.03 | 8.46 | 10.24 | 11.84 |
| 15km+2km | 2.06 | 3.68 | 5.16 | 6.90 | 7.92 | 8.55 |
| 20km+2km | 2.24 | 3.41 | 4.15 | 6.49 | 7.22 | 8.22 |
| 12km+12km | 3.86 | 6.40 | 11.08 | 24.04 | 24.04 | 34.73 |

Table II displays the QKD index using the new integration method when both QKD and GPON are working at the same time.

Table II.. Key generation rate of different fiber lengths and different optical splitting ratios.

| Fiber Length | Optical splitting Ratio | | | | | |
|---|---|---|---|---|---|---|
| | 1:4 | 1:8 | 1:16 | 1:32 | 1:64 | 1:128 |
| 12km+2km | 10900 bps | 10600 bps | 10400 bps | 10300 bps | 35000 bps | 0 |
| 15km+2km | 5130 bps | 4400 bps | 4300 bps | 0 | 0 | 0 |
| 20km+2km | 4200 bps | 3700 bps | 1900 bps | 0 | 0 | 0 |
| 12km+12km | 2700 bps | 0 | 0 | 0 | 0 | 0 |

A high QKD bit rate can be obtained using the new method, when quantum and classical optical signals can be transmitted in the same fiber. Neither GPON nor QKD can work using a traditional integrating method. Note that both QKD and GPON can run and that a high QKD bit rate can be obtained when the fiber length is 12km+2km and the optical splitting ratio is 1:64. A ratio of 1:64 is most commonly used. Both QKD and GPON can run and a high QKD bit rate can be obtained with an optical splitting ratio of 1:16 and a fiber length of 15km+2km. We set the upper quantum bit error rate (QBER) to 3% to get a sizable key rate. However, theoretically the QKD system can tolerate at most 11% QBER[36]. A secure key using privacy amplification can be obtained even when QBER is 6%. So, if we set higher rates, QKD and GPON both can run with a longer fiber length and a larger optical splitting ratio.

We have proposed an integration program of QKD and GPON based on WDM. The quantum signals are cleaned and multiplexed at the optical splitting point. The quantum signals bypass the optical splitter, avoiding losses produced by the optical splitters. The counting rate of the signal state increases and the SNR improved, so the QKD can obtain a higher key generation rate and a longer transmission distance. Also, we can enlarge the key generation rate and transmission distance by improving the upper value of the acceptable QBER and using a fiber Bragg grating as a denser filter. WDM, QKD and GPON enable us to transmit quantum signal using existing fiber infrastructure, reducing the costs of practical QKD and setting a stable foundation for QKD to be used by ordinary users.

This work has been supported by the Excellent Young Talents Fund Program of Higher Education Institutions of Anhui Province and the Fundamental Research Funds for the Central Universities. The authors would like to thank QuantumCTek Co., Ltd., for providing the QKD system.